\documentclass[reprint,pra, longbibliography,superscriptaddress,amsmath,amssymb,aps]{revtex4-2}

\usepackage{graphicx}
\usepackage{dcolumn}
\usepackage{bm}
\usepackage[T1]{fontenc}
\usepackage{siunitx}
\usepackage{chemformula}
\usepackage{CJKutf8}
\usepackage{xcolor}
\usepackage{hyperref}

\newcommand{\pd}[2]{\frac{\partial #1}{\partial #2}} 

\newboolean{ShowComments}
\setboolean{ShowComments}{true}  
\ifthenelse{\boolean{ShowComments}}%
	{
		\newcommand{\ColorComment}[3]{%
				{\colorbox{#1}{\color{white}   \textsf{\textbf{#2}}} \textcolor{#1}{#3}}}
		\newcommand{\nyacite}[1]{[#1]}
	}%
	{
		\newcommand{\ColorComment}[3]{}
		\newcommand{\nyacite}[1]{}
	}%
\definecolor{newcolor}{rgb}{0.5,0,0}

\usepackage{cancel}

\begin{document}

\title{Depth-Resolved Lattice Distortions in a Silicon-Germanium Qubit Host}
\author{Jonathan C. Marcks}
\altaffiliation{Corresponding author: Jonathan C. Marcks, jmarcks@anl.gov}
\affiliation{Q-NEXT, Argonne National Laboratory, Lemont, Illinois, 60439, United States}
\affiliation{Materials Science Division, Argonne National Laboratory, Lemont, Illinois, 60439, United States}
\affiliation{Pritzker School of Molecular Engineering, University of Chicago, Chicago, Illinois, 60637, United States}

\author{E. S. Joseph}
\affiliation{Department of Physics, University of Wisconsin-Madison, Madison, Wisconsin, 53706, United States}
\affiliation{Q-NEXT, Argonne National Laboratory, Lemont, Illinois, 60439, United States}

\author{J. Reily}
\affiliation{Department of Physics, University of Wisconsin-Madison, Madison, Wisconsin, 53706, United States}
\affiliation{Q-NEXT, Argonne National Laboratory, Lemont, Illinois, 60439, United States}

\author{Talise Oh}
\affiliation{Department of Physics, University of Wisconsin-Madison, Madison, Wisconsin, 53706, United States}

\author{Abigail Postlewaite}
\affiliation{Department of Physics, University of Wisconsin-Madison, Madison, Wisconsin, 53706, United States}

\author{Tao Zhou}
\affiliation{Center for Nanoscale Materials, Argonne National Laboratory, Lemont, Illinois, 60439, United States}

\author{M. A. Eriksson}
\affiliation{Department of Physics, University of Wisconsin-Madison, Madison, Wisconsin, 53706, United States}

\author{Mark Friesen}
\affiliation{Department of Physics, University of Wisconsin-Madison, Madison, Wisconsin, 53706, United States}

\author{Martin V. Holt}
\affiliation{Q-NEXT, Argonne National Laboratory, Lemont, Illinois, 60439, United States}
\affiliation{Center for Nanoscale Materials, Argonne National Laboratory, Lemont, Illinois, 60439, United States}

\begin{abstract}
Semiconductor qubits, promising for quantum computation, inherit properties from their host lattice. Quantum dot spins, occupying the local lowest energy states in the conduction band, necessarily couple to structural disorder and interfaces. While silicon-based systems promise low noise alongside industrially compatible manufacturing, the growth of SiGe---a leading platform---unavoidably introduces lattice dislocations, inhomogeneous strain, and crosshatch patterns, expected to cause fluctuations between devices, qubit failure, and subsequently higher operational overhead. Through X-ray nano-structural mapping of an Intel Si/SiGe chip, we reveal, with \SI{30}{\nano\meter} lateral and \SI{200}{\nano\meter} functional depth resolution, how extended lattice defects introduced during growth propagate through the heterostructure, creating permanently distorted lattice planes and strain. We correlate these at the $\approx\SI{1}{\micro\meter}$ scale of a quantum dot device and calculate the impact on qubit energy spectra. We observe crosshatch fine structure and find that substrate miscut and growth correlate with the final crosshatch pattern.
\end{abstract}

\maketitle

Semiconductors host spatially localized spin qubits in many emerging quantum devices~\cite{burkard_semiconductor_2023,chatterjee_semiconductor_2021,scappucci_germanium_2021,vandersypen_interfacing_2017,wolfowicz_quantum_2021,deleon2021,awschalom2018,marcks_nuclear_2025}. Silicon-based quantum dot (QD) systems in particular promise to combine decades of fabrication expertise with material properties such as a nuclear spin-free isotope and low spin-orbit coupling to realize low-noise, scalable qubits with sub-micron footprints~\cite{eriksson_spin-based_2004,zwerver_qubits_2022,neyens_probing_2024,george_12-spin-qubit_2025,steinacker_industry-compatible_2025,elsayed_low_2024,tomic_long_2025,beyne_300mm_2025,koch_industrial_2025,madzik_operating_2025}. Gate-defined QD spin qubits, which rely on transistor-like fabrication on heterostructures to electrostatically trap single electrons, have demonstrated fault tolerant-compatible fidelities~\cite{noiri_fast_2022,blumoff_fast_2022,wu_simultaneous_2025} in industrially manufactured devices with multiple qubit encodings~\cite{steinacker_industry-compatible_2025,team_digitally_2026}. These systems are formed in a two-dimensional electron gas at the conduction band minimum of the host material, such that qubit energy levels are directly determined by the local band structure of the host, and are thus sensitive to the local crystal structure~\cite{burkard_semiconductor_2023,marcks_valley_2025,losert_practical_2023,dodson_how_2022,volmer_impact_2026,volmer_mapping_2026,klos_atomistic_2024,volmer_mapping_2024,paquelet_wuetz_atomic_2022,pena_cross-hatch_2026}.

Strained silicon quantum wells in alloyed silicon germanium ($\epsilon$Si/SiGe) offer the high electron mobility and low spin noise suitable for forming QD spin qubits~\cite{schaffler_high-mobility_1997,eriksson_spin-based_2004}. However, the same lattice constant mismatch that contributes to a deep potential well in the $\epsilon$Si, where the qubit resides, also poses challenges for the integration of SiGe with Si substrates necessary to produce devices. Epitaxy of SiGe on Si necessarily introduces dislocations to alleviate the lattice mismatch~\cite{mooney_strain_1996,schaffler_high-mobility_1997,sawano_-plane_2003}. While current approaches partly mitigate dislocations through graded SiGe buffer layers---where Ge is increasingly introduced throughout microns of growth---the heterostructures nonetheless present a so-called crosshatch pattern (CHP): buckling of the lattice planes, observed as surface height variation or lattice plane tilt, resulting from dislocation bundles forming along the SiGe \{111\} slip planes~\cite{fitzgerald_totally_1991,degli_esposti_low_2024,pena_cross-hatch_2026}. The CHP leads to inhomogeneity in strain and lattice tilt, which distorts the crystal and thus the emergent QD qubits~\cite{degli_esposti_low_2024,pena_cross-hatch_2026}.

Nanoscale X-ray diffraction microscopy (nXDM)~\cite{winarski_hard_2012,holt_nanoscale_2013} provides a tool to measure the lattice distortions inside these heterostructure hosts at the lengthscale and sensitivity relevant to quantum devices. Previous nXDM SiGe quantum dot heterostructure work has studied lattice tilt in the quantum well~\cite{evans_nanoscale_2012}, electrode-induced stress and curvature~\cite{park_electrode-stress-induced_2016}, composition heterogeneity~\cite{zoellner_imaging_2015}, and the full strain tensor~\cite{corley-wiciak_nanoscale_2023,corley-wiciak_lattice_2023}. These complement atomic force and Raman microscopies~\cite{perova_composition_2011,rovaris_dynamics_2019,degli_esposti_low_2024,pena_cross-hatch_2026}, as well as single-qubit spectroscopy~\cite{chen_detuning_2021,marcks_valley_2025} that probes low-energy excitations. In this work, we image an industrially manufactured Si/SiGe device from Intel (Tunnel Falls) with sub-micron lateral (\SI{30}{\nano\meter}) and, for the first time to our knowledge, depth (\SI{200}{\nano\meter}, extracted through analysis) spatial resolution. We reveal how lattice strain and CHP tilt arise during heterostructure growth, correlate with one another, and impact the $\epsilon$Si quantum well (QW) qubit host. We first describe the nXDM experiment, focusing on the depth-resolved diffraction enabled by the graded SiGe buffer layer. We then use this functional depth resolution to study the emergence of CHP through the graded buffer. Through high-resolution imaging around a qubit device we correlate strain and CHP, as well as reveal fine structure in the CHP. Finally, we present calculations quantitatively relating the measured lattice strain and tilt to properties of the quantum well---energy fluctuations and atomic step density---that are expected to affect qubit operation. Taken together, this study links qubit host growth with qubit performance in order to guide future development of heterostructures and devices.

\begin{figure}
    \centering
    \includegraphics{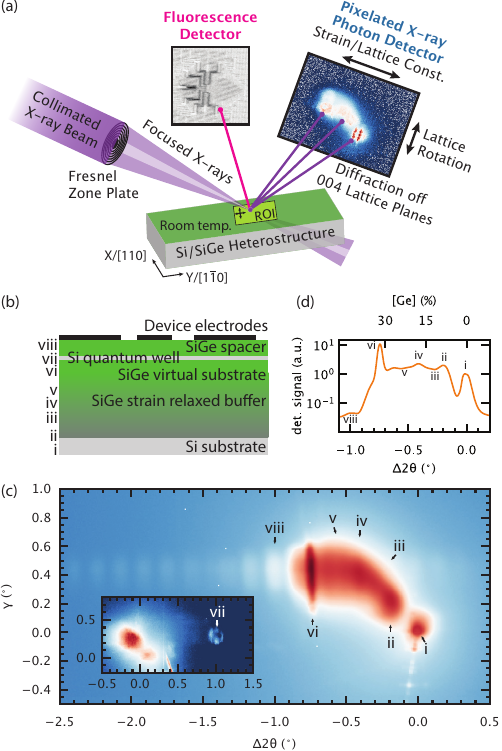}
    \caption{(a) nanoscale X-ray diffraction microscopy (nXDM) of the 004 peak of the Si/SiGe heterstructure shown in (b), where the Pixelated X-ray Photon Detector maps the crystal diffraction and the Fluorescence Detector maps material composition. Detector images show example diffraction signal of the heterostructure and fluorescence from the qubit device electrodes. The scanning $X$ (transverse to beam path) and $Y$ directions are $\approx\SI{3}{^\circ}$ to the $[110]$ and $[1\bar{1}0]$ crystal axes, due to sample mounting. All measurements are performed at room temperature. (b) Schematic of SiGe/Si/SiGe/Si heterostructure under study (not to scale). Labels at left correspond to features in diffraction (c,d). (c) Parallel beam diffraction signal with labeled features: (i) Si substrate, (ii) initial jump in growth to \ch{Si_{.93}Ge_{.07}}, (iii) graded buffer up to \ch{Si_{.81}Ge_{.19}}, (iv) kink at \ch{Si_{.81}Ge_{.19}}, where overall surface tilt is fixed, (v) graded buffer up to \ch{Si_{.69}Ge_{.31}}, (vi) \ch{Si_{.69}Ge_{.31}} virtual substrate, (viii) \ch{Si_{.69}Ge_{.31}} spacer atop quantum well. Inset: Focused beam diffraction at longer exposure, showing (vii) Si quantum well at expected tilt relative to the SiGe virtual substrate. (d) Summed detector signal showing correspondence between $\Delta2\theta$ and [Ge] in the graded buffer, calculated assuming diffraction shift arises from the SiGe lattice constant and that the Si substrate is unstrained.}
    \label{fig:fig1}
\end{figure}

We employ the X-ray nanoprobe depicted in Fig.~\ref{fig:fig1}(a) to study the Si/SiGe heterostructure shown schematically in Fig.~\ref{fig:fig1}(b). nXDM is performed at room temperature by focusing a collimated \SI{9.5}{\kilo\eV} X-ray beam to a \SI{30}{\nano\meter} spot size at the Si/SiGe 004 Bragg condition. As this beam is scanned around the region of interest, a Pixelated X-ray Photon Detector collects the entire diffraction pattern with \SI{0.009}{^\circ} angular resolution, as in the inset, where the characteristic donut shapes result from the Fresnel Zone Plate. The diffraction signal maps out the local lattice vector in the sample---the lattice constant, strain, and lattice rotation all deflect the outgoing X-ray beam and are thus captured in the diffraction signal. The Fluorescence Detector collects X-ray fluorescence, providing a spatial map of patterned features. The narrowly focused beam has a depth of focus much larger than the heterostructure, such that the diffraction volume encompasses the full depth of the heterostructure.

The heterostructure, produced by Intel for Tunnel Falls spin qubit processors, is designed to host a two-dimensional electron gas, and subsequently single electron quantum dots, in the quantum well formed by sandwiching a few-nanometer layer of $\epsilon$Si (vii, referencing the labels throughout Fig.~\ref{fig:fig1}) between two layers of SiGe (vi, viii), inducing a large band offset in the Si~\cite{schaffler_high-mobility_1997}. Integrating high-Ge content SiGe ($\approx 30\%$) with high-quality Si substrate wafers (i) requires the growth of a strain-relaxed buffer layer (ii-v) that increments the Ge concentration up to the desired concentration. Here, in the Tunnel Falls device, [Ge] increases linearly up to the so-called virtual substrate (vi)~\cite{george_12-spin-qubit_2025}. This layer is necessary to mitigate the strain and dislocations that arise from the lattice mismatch between Si and SiGe. However, as we show here, these features propagate through the SiGe growth and are present in the quantum well.

Fig.~\ref{fig:fig1}(c) shows the parallel beam diffraction signal of the Si/SiGe. The axes are referenced to the Si substrate 004 peak assuming the substrate is unstrained. Corresponding to the schematic in Fig.~\ref{fig:fig1}(b), we identify distinct signals from the Si substrate through the buffer layer, virtual substrate, and \SI{50}{\nano\meter} SiGe cap layer, which is visible as a series of interference peaks. The signal from the thin quantum well (QW) is not visible here. Inset in Fig.~\ref{fig:fig1}(c) is a long-exposure focused beam diffraction taken at larger incident angle where a component of the QW diffraction is visible around $\Delta2\theta=\SI{1}{^\circ}$. The total diffraction from the thin QW is spread out in $\Delta2\theta$, but is at comparable $\gamma$ to the virtual substrate signal. In the Supplemental we extract the tilt of the QW along the studied qubit device and find it agrees with the tilt observed in the virtual substrate in Fig.~\ref{fig:fig4}(c), confirming that the QW is epitaxial with the virtual substrate and justifying our assumption throughout this paper that the strain and lattice tilt measured in the virtual substrate are also present in the QW. Strain and lattice tilt throughout this work are extracted by defining ``virtual detectors'' over regions of interest in the diffraction signals of scanning tilt series, where the sample angle is rocked, described in the Supplemental.

The lattice constant-dependent diffraction combined with the [Ge]-dependent lattice constant enables functionally depth-resolved diffraction, circumventing the restriction of the large X-ray penetration depth, demonstrated in Fig.~\ref{fig:fig1}(d). Fig~\ref{fig:fig1}(d) is the detector signal summed in $\gamma$ with heterostructure features identified. The top axis shows the [Ge] at each $\Delta2\theta$ angle, calculated from the lattice constant by inverting $a_{\mathrm{Si}_{1-x}\mathrm{Ge}_x}=0.027x^2+0.2x+a_{\mathrm{Si}}$~\cite{dismukes_thermal_1964}. We ascribe the diffraction signal at a given $\Delta 2\theta$ to a depth in the graded buffer layer by comparing \ref{fig:fig1}(e) to an independent measurement of [Ge] versus depth via secondary ion mass spectrometry (SIMS) for a similarly grown heterostructure.

\begin{figure}
    \centering
    \includegraphics{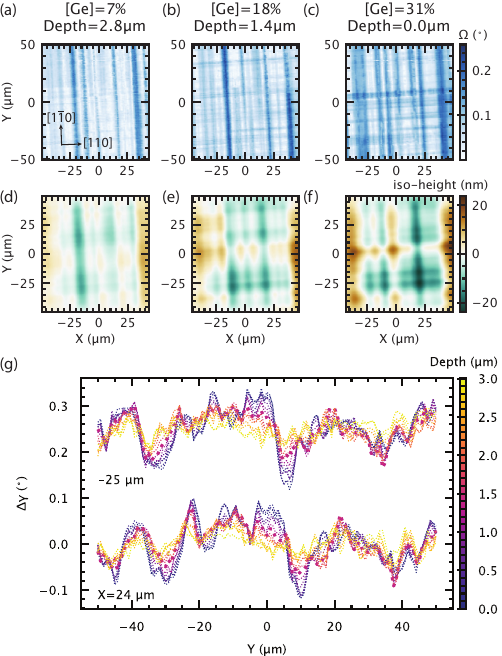}
    \caption{Lattice tilt, $\Omega=\sqrt{\gamma^2+\theta^2}$, maps at the start of the buffer layer (a), partway through the buffer layer (b), and at the virtual substrate (c), calculated from five pixel-wide virtual detectors in $\Delta2\theta$, with (d-f) corresponding lattice iso-surface heights calculated by integrating the tilt. Scale bars are shared in each row. (g) Line-cuts of $\gamma$ (horizontal striations in $\Omega$) at all depths and two $X$ values, with the \SI{1.2}{\micro\meter} depth, corresponding to feature iv in Fig.~\ref{fig:fig1}(c), bolded.}
    \label{fig:fig2}
\end{figure}

The functionally depth-resolved diffraction allows the observation of lattice tilt throughout the graded buffer layer, providing insight into how substrate preparation and treatment impact tilt, and thus, as we demonstrate below in Fig.~\ref{fig:fig3}, device strain. To spatially filter diffraction from a given lattice constant we form 5~pixel-wide virtual detectors in $\Delta2\theta$ and extract $\theta$ and $\gamma$ angular tilts from the signal center-of-mass motion, resulting in $\approx\SI{200}{\nano\meter}$ depth resolution, calculated from the associated [Ge]. $\gamma$ corresponds to tilt about the $[110]$ axis, oriented at a slight angle to the $X$ scanning axis, and $\theta$ corresponds to tilt about the $[1\bar{1}0]$ axis, oriented at a slight angle to the $Y$ scanning axis. We remove the arbitrary offset in each and then combine into $\Omega=\sqrt{\gamma^2+\theta^2}$, which captures the lattice plane slope regardless of direction. As $\gamma$ and $\theta$ are perpendicular, this effectively overlays the two into a single CHP. Fig.~\ref{fig:fig2}(a-c) show $\Omega$ tilt maps at three evenly spaced depths, referenced to the start of the virtual substrate and shown with corresponding [Ge]. Fig.~\ref{fig:fig2}(a) is the start of SiGe growth at [Ge]$=7\%$; $\theta$ tilt (vertical striations) is already visible, however there is minimal $\gamma$ tilt (horizontal striations). This is consistent with an initial substrate miscut along $[1\bar{1}0]$, which we also observe through the gradual increase in average $\gamma$ in the diffraction signals in Fig.~\ref{fig:fig1}. As the graded buffer grows, through Fig.~\ref{fig:fig2}(b) and (c), $\theta$ features remain relatively constant while $\gamma$ features appear and propagate up to the virtual substrate, forming the final CHP. As the lattice tilts are a direct measure of the slope of the lattice planes, we integrate the tilt maps to reveal the spatial undulations of the lattice isosurfaces, shown in Fig.~\ref{fig:fig2}(d-f). New variations in the lattice isosurfaces appear as new tilt features arise.

In Fig.~\ref{fig:fig2}(g) we show line-cuts of $\Delta\gamma=\gamma - \overline\gamma$ at two values of $X$, with the line-cuts at a depth of \SI{1.2}{\micro\meter} bolded. This depth corresponds to the kink in the diffraction signal related to the surface tilt (Fig.~\ref{fig:fig1}(c), feature iv). We observe at multiple locations, near $X=$\SI{-45}{\micro\meter}, \SI{-30}{\micro\meter}, \SI{0}{\micro\meter}, and \SI{10}{\micro\meter}, a new tilt feature appears at this depth and propagates up to the virtual substrate, indicating a correlation between strain relaxation in the graded buffer and structural distortions in the virtual substrate.

\begin{figure}
    \centering
    \includegraphics{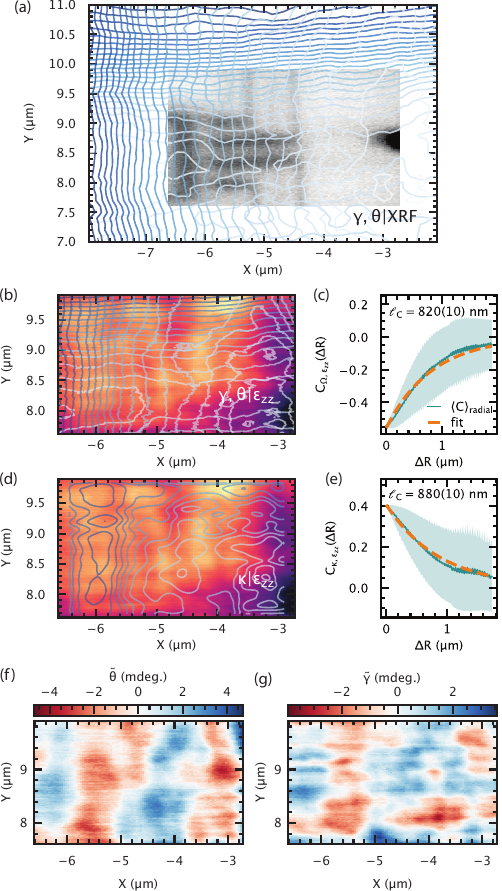}
    \caption{(a) $\gamma$ and $\theta$ tilt contours overlaid on X-ray fluorescence (XRF) of device electrodes, which define the active qubit region. (b,d) $\gamma$ and $\theta$ tilt contours (b) and $\kappa$ curvature contours (d) overlaid on $c$-axis strain $\epsilon_{zz}$. (c,e) Radially averaged cross-correlation $C_{\Omega,\epsilon_{zz}}$ (c) and $C_{\kappa,\epsilon_{zz}}$ (e) as a function of displacement $\Delta R$, fit empirically to a mono-exponential decay with decay length $\ell_C=\SIrange{820}{880}{\nano\meter}$. Shaded areas show one standard deviation from the radial averaging. (f,g) Fine-structure tilt maps $\tilde\theta$ and $\tilde\gamma$ after second-order polynomial background subtraction.}
    \label{fig:fig3}
\end{figure}

In Fig.~\ref{fig:fig3} we further investigate the correlation between the lattice tilt $\Omega$ and curvature $\kappa$ and $c$-axis strain $\epsilon_{zz}$, where $\kappa=\left|\pd{[\theta,\gamma]}{[X,Y]}\right|$ is the Jacobian of the tilt.
Fig.~\ref{fig:fig3}(a) shows the lattice tilt contours measured in the virtual substrate overlaid on the X-ray fluorescence focused on a qubit device, with confining gate electrodes visible, which shape the 2D electron gas into qubit and sensor channels. The device is off-center in the image to ensure that the angled X-ray beam samples the virtual substrate, and thus can measure the strain, under the device. This particular device region sits at the corner of two intersecting perpendicular crosshatch features. We focus on this region to study strain-tilt correlation at a device-scale without interference of the large-scale crosshatch.

Fig.~\ref{fig:fig3}(b) overlays contours of $\gamma$ and $\theta$ atop a $\epsilon_{zz}$ heatmap. Strain and tilt are studied quantitatively in Fig.~\ref{fig:fig4} and Fig.~\ref{fig:fig5}; here we focus on spatial cross-correlations. Fig.~\ref{fig:fig3}(c) shows the cross-correlation $C_{\Omega,\epsilon_{zz}}(\Delta R)$ as a function of spatial displacement $\Delta R$, radially averaged over all angles out from zero displacement. Empirically fitting to a mono-exponential decay reveals a correlation decay length $\ell_C$=\SI{820\pm10}{\nano\meter} with a maximum negative correlation of $-0.55$, showing that larger (smaller) lattice tilt is correlated with more negative, larger amplitude (less negative, smaller magnitude) strain. Performing the same analysis for curvature $\kappa$ in Fig.~\ref{fig:fig3}(d) and (e) reveals a similar correlation decay length of $\ell_C$=\SI{880\pm10}{\nano\meter} and maximum positive correlation of $0.41$. The existence of correlations with both $\Omega$ and $\kappa$ confirm that CHP features, directly observed through the lattice distortion, have high curvature and tend to bundle strain, allowing for low-tilt, small-strain regions of the material.

The nanoscale spatial resolution reveals that these features are only correlated at $<\SI{1}{\micro\meter}$. At this scale, in addition to the crosshatch tilt, there is finer tilt structure visible along the crosshatch slopes, shown as $\tilde\theta$ and $\tilde\gamma$ in Fig.~\ref{fig:fig3}(f) and (g), respectively, by background subtracting a second-order polynomial from the device-scale tilt maps. This reveals features that otherwise are drowned out by the CHP, but likely contribute to strain relaxation nonetheless. The largest fine-structure tilt of \SI{0.0046}{^\circ} is a slope of \SI{0.8}{\angstrom}/\SI{1}{\micro\meter}.

\begin{figure}
    \centering
    \includegraphics{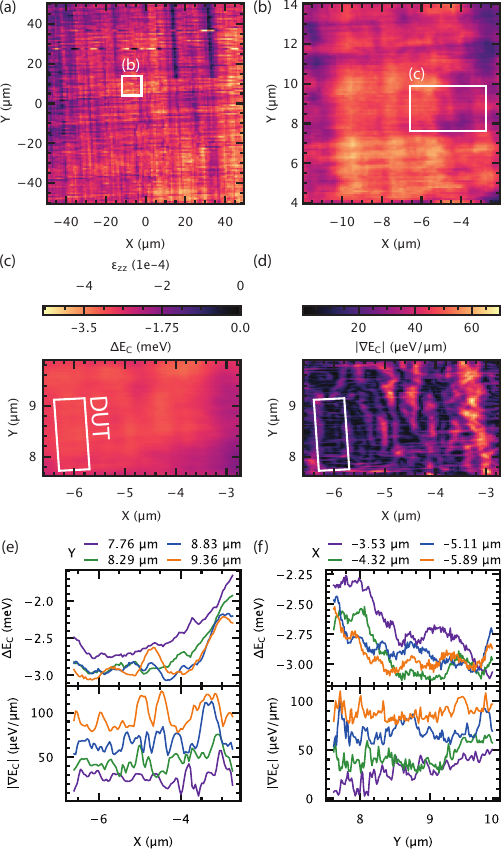}
    \caption{(a-c) c-axis strain $\epsilon_{zz}$ and proportional conduction band minimum shift $\Delta E_c$ at three lengthscales surrounding the qubit device (DUT). White boxes show zoomed-in regions of subsequent figures and DUT. Measurement artifacts are visible from beam drops between $Y=\SIrange{15}{50}{\micro\meter}$ in (a). The scale bar in (c) is shared with (a) and (b). (d) Magnitude of the gradient of $\Delta E_C$, $|\nabla E_C|$, zoomed in on the device. The spatial sampling ($\Delta X=\SI{39}{\nano\meter}$, $\Delta Y=\SI{13}{\nano\meter}$) in this image is small enough to fully sample the strain. (e,f) Line-cuts at various $Y$ ($X$) values of $\Delta E_C$ and $|\nabla E_C|$. $|\nabla E_C|$ plots are offset by \SI{25}{\micro eV/\micro\meter} for visibility.}
    \label{fig:fig4}
\end{figure}

We now study the impact of strain on the emergent energy landscape of QD spin qubits. In Fig.~\ref{fig:fig4}(a-c) we show the $c$-axis strain $\epsilon_{zz}$ in the virtual substrate, and thus the QW, and the resulting proportional shift to the conduction band minimum energy $\Delta E_C$, at three lengthscales of $(\SI{100}{\micro\meter})^2$ (a), $(\SI{10}{\micro\meter})^2$ (b), and $\SI{4}{\micro\meter}\times\SI{2}{\micro\meter}$ (c), focused on the qubit device. The colorbar in (c) is shared with (a,b). The labeled DUT refers to the qubit channel within the device shown in the Fluorescence Detector in Fig.~\ref{fig:fig1}(b). $\Delta E_C$ is calculated as $\xi_u\epsilon_{zz}+\xi_d(2\epsilon_{xx}+\epsilon_{zz})$, where $\epsilon_{xx}=\epsilon_{zz}\cdot\frac{s_{11}+s_{12}}{2s_{31}}$, $s_{ij}$ are components of the stiffness tensor, and $\xi_u,~\xi_d$ are deformation potentials~\cite{sverdlov_strain-induced_2011}. Strain down to \SI{-5e-4}{} is present across the imaged region, with fine structure at both the $\mu$m and sub-$\mu$m scales. Strain in the QW directly affects the conduction band offset in the well, introducing shifts in a QD chemical potential, energy detuning between neighboring dots, and long-range energy fluctuations. These may affect qubit operation, in particular double-dot operations like exchange coupling, and electron shuttling, where electrons move over microns inside the QW. To further study this, we plot the gradient of the energy shift $|\nabla E_C|$ in Fig.~\ref{fig:fig4}(d), with linecuts of $\Delta E_C$ and $\nabla E_C$ in Fig.~\ref{fig:fig4}(e,f). While the gradient over the studied DUT is by chance low, we observe areas as high as \SI{68}{\micro \eV/\micro\meter}. As valley splitting energies in SiGe can be $\approx\SI{100}{\micro\eV}$~\cite{dodson_how_2022,marcks_valley_2025,volmer_mapping_2024}, this gradient corresponds to a change equal to the valley splitting energy over the length of a moderate-size device, and may impact shuttling fidelity~\cite{losert_strategies_2024}.

\begin{figure}
    \centering
    \includegraphics{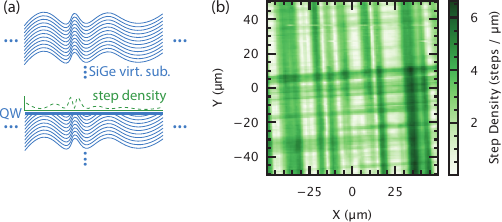}
    \caption{(a) As the strain relaxed buffer and virtual substrate grow, lattice tilt buckles the lattice planes (top). After polishing and re-growth, the quantum well intersects the truncated lattice planes, leading to finite step density at the well interface (bottom). (b) Atomic step density calculated from lattice tilt $\Omega$.}
    \label{fig:fig5}
\end{figure}

In Fig.~\ref{fig:fig5} we study how lattice tilt results in atomic steps at the quantum well interface. Fig.~\ref{fig:fig5}(a) shows schematically how atomic steps arise even with polishing during heterostructure growth. On the top is a set of lattice planes that are distorted by spatially varying lattice tilt. On the bottom, even if the surface is polished, the lattice tilt remains in place. When the Si quantum well is grown epitaxially atop the distorted SiGe lattice planes, the truncated planes intersect the well interface, causing atomic steps, plotted schematically above the quantum well. Fig.~\ref{fig:fig5}(b) calculates the step density $\tan(\Omega)/a$ from the tilt map. Regions where tilt is higher consequently have higher step density, up to \SI{6.6}{steps/\micro\meter}. At the scale of around 10 quantum dots, this corresponds to multiple dots intersected by atomic steps, which is known to significantly reduce valley splitting in the deterministic regime by a factor of 3-4~\cite{friesen_magnetic_2006,friesen_valley_2007,losert_practical_2023}, although this is dependent on the orientation of the device.

We have demonstrated that spatially varying strain and lattice tilt, directly measured in nXDM, have the potential to impact the behavior of scaled-up quantum dot processors through their coupling to the qubit energy landscape. In the Supplemental we explore if strain and lattice tilt correlate in this specific device with valley splitting energy, characterized in Ref.~\cite{marcks_valley_2025}. We previously determined that the distribution of valley splitting was properly described by the alloy disorder in the engineered quantum well, and here do not find any relationship to the measured lattice distortions. However, this device was not designed to be specifically sensitive to strain and tilt, and there are multiple factors that could lead to our observations: the device sits in a relatively flat strain region; alloy disorder is large; and the quantum well interface is broad, obscuring effects of atomic steps. Ultimately, fully correlating heterostructure features with spin qubit properties will require cryogenic qubit characterization with scale and resolution comparable to nXDM, as in recent shuttling experiments~\cite{volmer_mapping_2024,volmer_mapping_2026,volmer_impact_2026,de_smet_high-fidelity_2025,zwerver_shuttling_2023}.

Through functionally depth-resolved nXDM, sub-micron device-scale strain mapping, and theoretical calculations, we have shown how Si/SiGe heterostructure preparation and growth are expected to impact industrially manufactured qubit devices. The observed correlations between strain and tilt, consistent with the expected behavior of dislocation bundles, directly connect the evolution of lattice tilt to qubit energy fluctuations. While individual few-qubit devices can be situated in a low-strain, dislocation-free region of material, as devices scale to larger numbers and integrate components such as electron shuttlers, avoiding inhomogeneity in the host material will be unavoidable. This may motivate designing devices to avoid crosshatch features, or designing new growth approaches that guide dislocation bundles. Our measurements also provide a quantitative dataset on the heterostructure, at multiple scales, that will inform future device theory and modeling.

\section{Acknowledgments}
The authors thank Nathaniel C. Bishop, Joelle Corrigan, Eric Henry, and Nazar Delegan for useful discussion. Work was supported by the U.S. Department of Energy, Office of Science, National Quantum Information Science Research Centers as part of the Q-NEXT center. T.O. was supported by Netherlands Ministry of Defence under Award No. QuBits R23/009. Work performed at the Center for Nanoscale Materials and Advanced Photon Source on APS beam time award (DOI: \href{https://doi.org/10.46936/APS-191441/60015139}{10.46936/APS-191441/60015139}), both U.S. Department of Energy Office of Science User Facilities, was supported by the U.S. DOE, Office of Basic Energy Sciences, under Contract No. DE-AC02-06CH11357. The authors acknowledge Intel Corp. for providing the sample and SIMS data.

\section{Author Contributions}
J.C.M., M.A.E., and M.V.H. conceived of the study; J.C.M., E.S.J., J.R., T.Z., and M.V.H. performed measurements and analyzed data; T.O., A.P., and M.F. provided theory support; J.C.M. coordinated the project and prepared the manuscript with input from all authors.

\section{Competing Interests}
The authors declare no competing interests.

\clearpage

\onecolumngrid

\setcounter{section}{0}
\renewcommand{\thesection}{S\arabic{section}}

\setcounter{figure}{0}
\renewcommand{\thefigure}{S\arabic{figure}}
\renewcommand{\figurename}{Figure} 

\setcounter{equation}{0}
\renewcommand{\theequation}{S\arabic{equation}}

\setcounter{table}{0}
\renewcommand{\thetable}{S\arabic{table}}

\section*{Supplementary Materials}

\section{Virtual detectors}
\begin{figure*}[h]
    \centering
    \includegraphics[width=\linewidth]{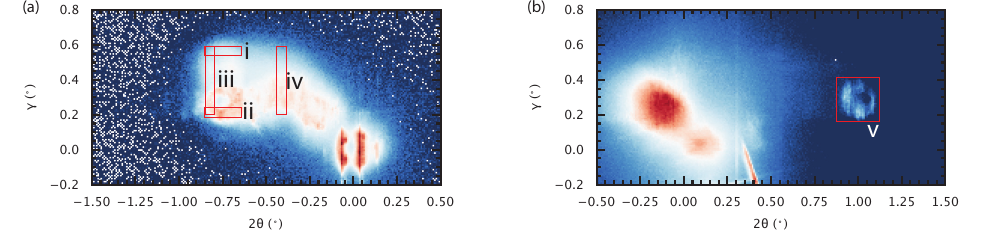}
    \caption{(a) Representative diffraction signal of heterostructure with virtual detector regions of interest defined. i and ii are sensitive to $\gamma$ tilt, while iii is sensitive to $\theta$ tilt. iv picks out a slice of the buffer layer, providing functional depth resolution. (b) Representative diffraction signal of quantum well, with virtual detector v defined around the QW diffraction feature.}
    \label{fig:figS1}
\end{figure*}

Two-dimensional diffraction maps are acquired for each spatial position and every angle during a scan, and contain signals from the entire illuminated volumn. In order to acquire signal related to a specific spatial region and lattice feature we track the center-of-mass (COM) signal within user-defined virtual detectors, shown in Fig.~\ref{fig:figS1}. Fig.~\ref{fig:figS1}(a) shows four representative virtual detectors overlaid on the diffraction signal from one of the spatial scans. We first focus on the right-most feature, around $2\theta=-0.75^\circ$, which corresponds to the virtual substrate. Multiple donut features are visible, arising from the spatially varying lattice tilt. By defining detectors i and ii we can track the $\gamma$ tilt, as the signal donut will shift in accordance with $\gamma$. Detector iii similarly tracks changes in $2\theta$. Detector iv shows how we isolate the signal from a specific depth in the graded buffer layer. By picking a window at a specific $2\theta$, we effectively filter the signal from a specific lattice constant and thus, as shown in main text Fig.~1, a specific depth. Fig.~\ref{fig:figS1}(b) reproduces the quantum well detector image from Fig.~1(d) with a virtual detector placed around the quantum well diffraction signal, used in Fig.~\ref{fig:figS2} to extract the quantum well tilt.

\section{Quantum well tilt}
\begin{figure}
    \centering
    \includegraphics[width=0.5\linewidth]{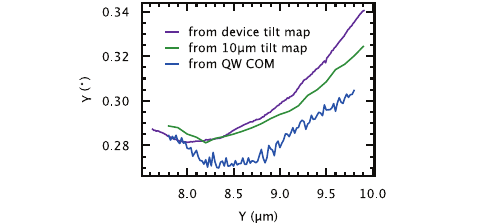}
    \caption{Comparison of $\gamma$ tilt along the qubit device extracted from the device-scale tilt map, the $(\SI{10}{\micro\meter})^2$ tilt map, and the quantum well diffraction signal (Fig.~\ref{fig:figS1}(b)).}
    \label{fig:figS2}
\end{figure}

In Fig.~\ref{fig:figS2} we compare the $\gamma$ tilt of the quantum well, extracted from the virtual detector COM in Fig.~\ref{fig:figS1}(b), with that of the device-scale and $(\SI{10}{\micro\meter})^2$ virtual substrate tilt maps in main text Fig.~4. We observe values in the virtual substrate tilt maps that match within 13\% of the QW values and exhibit the same structure over the QW is epitaxial with the virtual substrate. We do not expect perfect agreement because (1) the virtual substrate signal is averaged over $\approx\SI{1}{\micro\meter}$ of material, while the quantum well is only \SI{4.6}{\nano\meter}, and (2) the QW signal is affected by bleedthrough of background signal from the heterostructure diffraction, which can be seen in Fig.~\ref{fig:figS1}(b).

\section{Full depth dataset}
\begin{figure*}
    \centering
    \includegraphics[width=\linewidth]{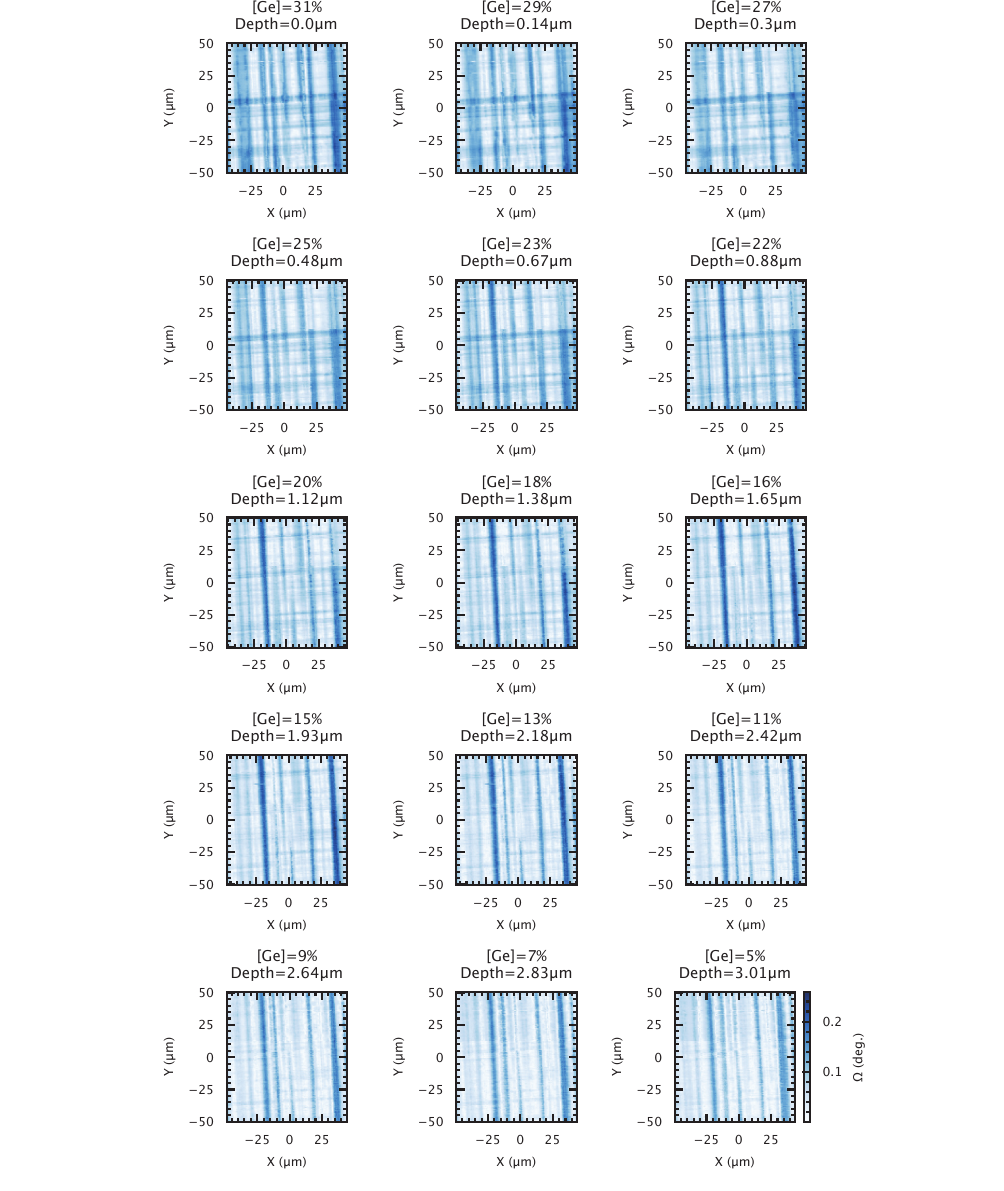}
    \caption{Full $\Omega$ tilt dataset for main text Fig.~2 for each depth and [Ge] extracted from the full heterostructure diffraction.}
    \label{fig:figS3}
\end{figure*}

\begin{figure*}
    \centering
    \includegraphics[width=\linewidth]{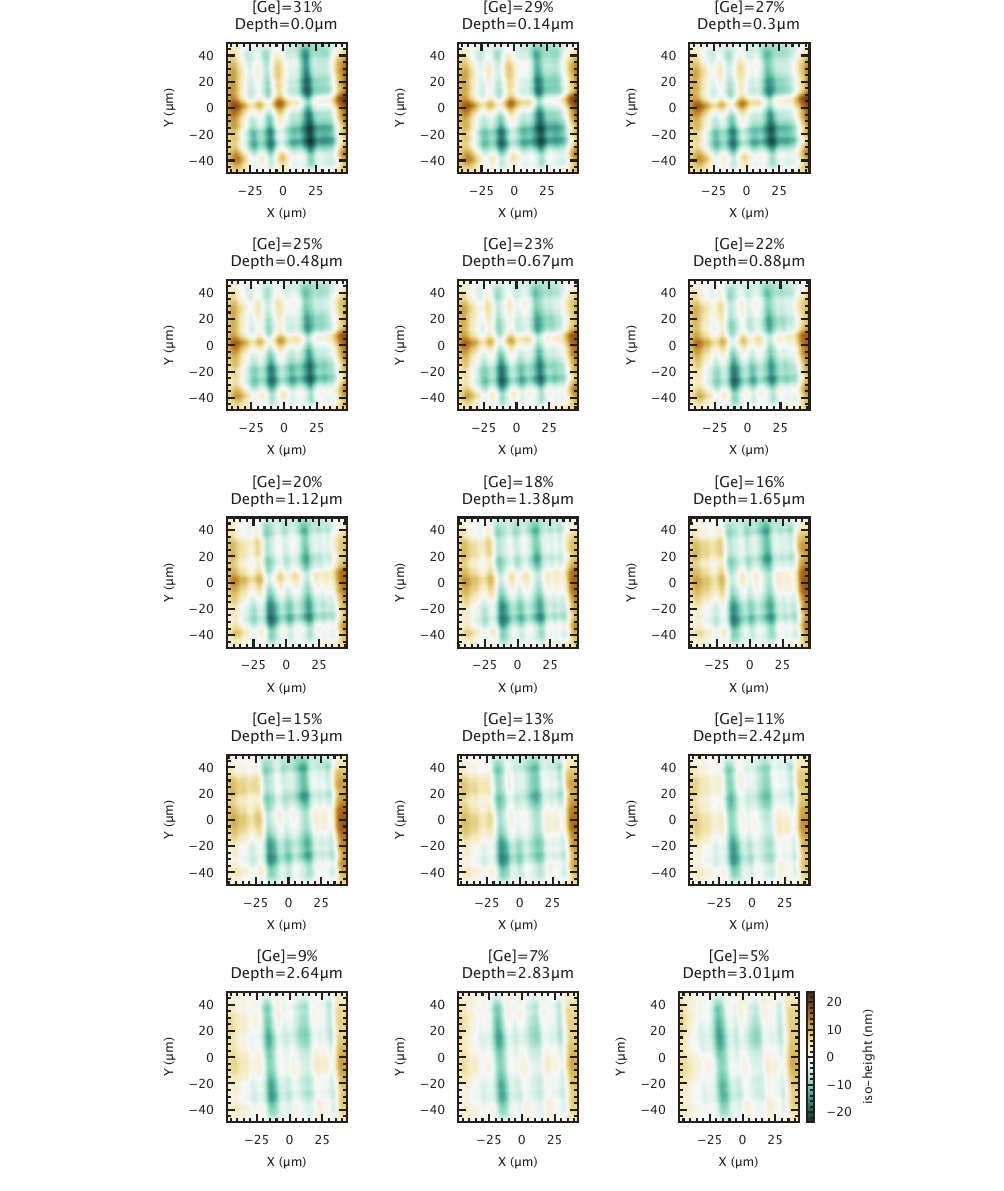}
    \caption{Full iso-surface dataset for main text Fig.~2 for each depth and [Ge] extracted from the full heterostructure diffraction.}
    \label{fig:figS4}
\end{figure*}

In Fig.~\ref{fig:figS3} and Fig.~\ref{fig:figS4} we show the tilt and iso-surface height for each analyzed depth in Fig.~2.

\section{Spatial cross-correlation}
\begin{figure*}
    \centering
    \includegraphics[width=\linewidth]{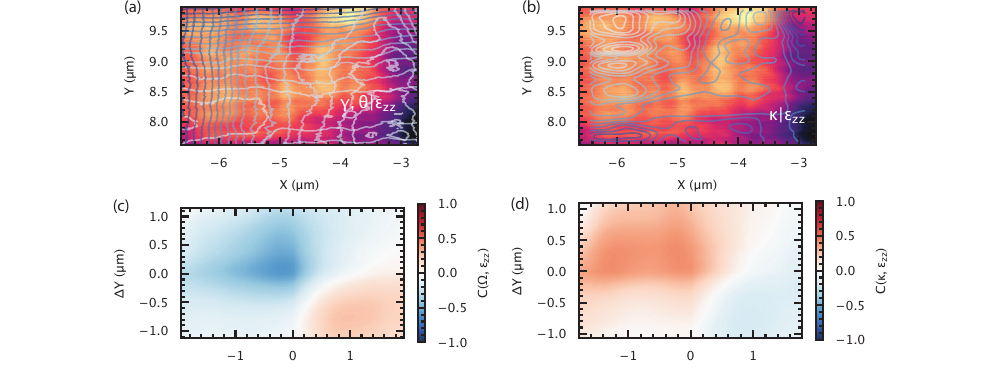}
    \caption{(a,b) $\gamma,\theta$ and $\kappa$ contours overlaid on $\epsilon_{zz}$ maps, reproduced from main text Fig.~3. (c,d) 2D spatial cross-correlation maps for $\Omega$ (c) and $\kappa$ (d), where correlation is calculated between the two 2D arrays for each displacement $(\Delta X,\Delta Y)$. Radially averaging over these maps produces the 1D cross-correlation in main text Fig.~3.}
    \label{fig:figS5}
\end{figure*}

In Fig.~3 we analyze the spatial cross-correlation between tilt $\Omega$, curvature $\kappa$, and strain $\epsilon_{zz}$. Spatial maps from Fig.~3 are reproduced in Fig.~\ref{fig:figS5}(a,b). For two spatial maps $A$ and $B$, we calculate the cross-correlation by centering about zero and normalizing each map
\begin{equation}
    \tilde A=\frac{A-\bar A}{\sqrt{\sum\left(A-\bar A\right)^2}}
\end{equation}
\begin{equation}
    \tilde B=\frac{B-\bar B}{\sqrt{\sum\left(B-\bar B\right)^2}}
\end{equation}
Then we calculate the cross-correlation $C$ with the SciPy function \begin{verbatim}C = scipy.signal.correlate2d(A,B)\end{verbatim}

\section{Comparison to valley splitting}
\begin{figure*}
    \centering
    \includegraphics[width=\linewidth]{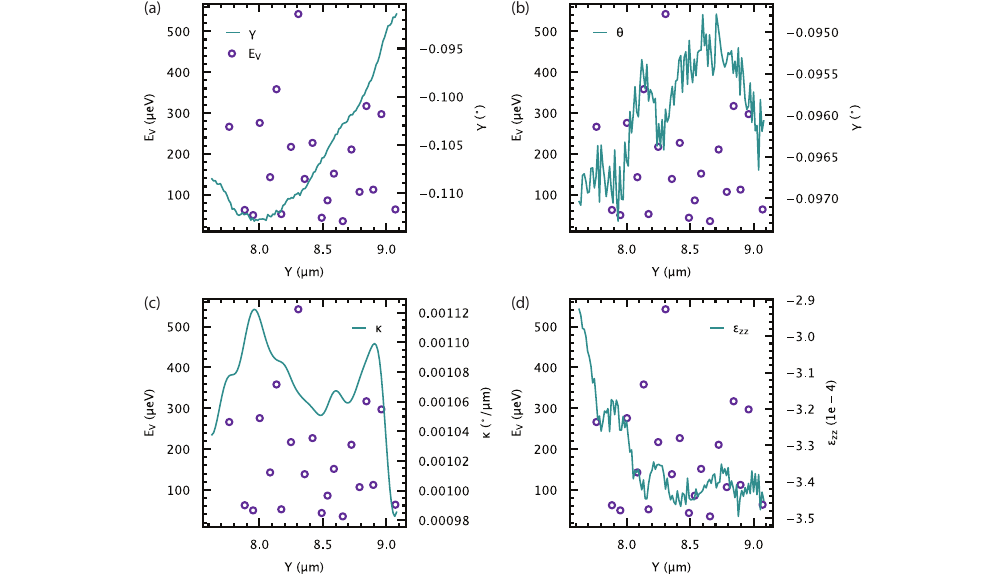}
    \caption{Comparison of valley splitting $E_{V}$ to lattice deformation: (a) $\gamma$, (b) $\theta$, (c) $\kappa$, and (d) $\epsilon_{zz}$.}
    \label{fig:figS6}
\end{figure*}

We have previously measured the valley splitting with \SI{60}{\nano\meter} resolution across the qubit channel of this device in Ref.~\cite{marcks_valley_2025}. We concluded that the observed wide and fast variation in valley splitting was completely explained by alloy disorder in the 2.8\%-Ge Si quantum well. However, there was a possibility that the lattice distortion, which we have now measured through X-ray diffraction, could explain the origin of the valley splitting, as in recent work~\cite{pena_cross-hatch_2026}. In Fig.~\ref{fig:figS6} we plot the previously measured valley splitting alongside the $\gamma$ tilt (a), $\theta$ tilt (b), curvature $\kappa$ (c), and $c$-axis strain $\epsilon_{zz}$ (d). No feature has the range or speed of variation to explain the dominant valley splitting fluctuations, supporting our initial conclusion about the role of alloy disorder.

\end{document}